\documentclass[paper]{JHEP3}
\usepackage{epsfig}
\catcode`@=11
\def\citer{\@ifnextchar
[{\@tempswatrue\@citexr}{\@tempswafalse\@citexr[]}}
\def\@citexr[#1]#2{\if@filesw\immediate\write\@auxout{\string\citation{#2}}\fi
  \def\@citea{}\@cite{\@for\@citeb:=#2\do
    {\@citea\def\@citea{--\penalty\@m}\@ifundefined
       {b@\@citeb}{{\bf ?}\@warning
       {Citation `\@citeb' on page \thepage \space undefined}}%
\hbox{\csname b@\@citeb\endcsname}}}{#1}}
\catcode`@=12

\def \ifb {fb$^{-1}$}
\def \gev {\mathrm{GeV}}

\def \lambdahhh {\ifmmode \lambda_{HHH}\else $\lambda_{HHH}$\fi}
\def \lambdahhhzero {\ifmmode \lambda_{HHH}^{(0)}\else
  $\lambda_{HHH}^{(0)}$\fi}
\def \lambdahhhh {\ifmmode \lambda_{HHHH}\else $\lambda_{HHHH}$\fi}
\def \lambdahhhhzero {\ifmmode \lambda_{HHHH}^{(0)}\else $\lambda_{HHHH}^{(0)}$\fi}
\def \mh {\ifmmode M_H \else $M_H$\fi}

\def    \nn             {\nonumber} 
\def    \=              {\;=\;} 
\def    \frac           #1#2{{#1 \over #2}}

\def    \lsim {\raisebox{-3pt}{$\>\stackrel{<}{\scriptstyle\sim}\>$}}  
\def    \gsim {\raisebox{-3pt}{$\>\stackrel{>}{\scriptstyle\sim}\>$}}

\newcommand     \be     {\begin{equation}} 
\newcommand     \ee     {\end{equation}} 
\newcommand     \ba     {\begin{eqnarray}} 
\newcommand     \ea     {\end{eqnarray}}

\newcommand     \ptmin     {\ifmmode p_{T}^{min} \else 
                            $p_{T}^{min}$ \fi}

\def \et   {\mbox{$E_{T}$}} 
\def \Emu  {\ifmmode{E_{\mu}
    }\else{$E_{\mu}$}\fi} 
\def \Enu  {\ifmmode{E_{\nu}
    }\else{$E_{\nu}$}\fi}

\def \nue  {\ifmmode{\nu_e}\else{$\nu_e$}\fi} 
\def \numu  {\ifmmode{\nu_{\mu}}\else{$\nu_{\mu}$}\fi}

\def \to   {\mbox{$\rightarrow$}}

\newcommand \jpsi{\ifmmode{J/\psi 
    }\else{$J/\psi$}\fi}

\def\Ord{\lower .7ex\hbox{$\;\stackrel{\textstyle <}{\sim}\;$}}
\def\OOrd{\lower .7ex\hbox{$\;\stackrel{\textstyle >}{\sim}\;$}}
\def\alpgen{{\small ALPGEN}}

\newcommand{\br}{\begin{eqnarray}}
\newcommand{\er}{\end{eqnarray}}
\newcommand{\bea}{\begin{eqnarray}}
\newcommand{\eea}{\end{eqnarray}}
\newcommand{\bi}{\begin{itemize}}
\newcommand{\ei}{\end{itemize}}
\newcommand{\bn}{\begin{enumerate}}
\newcommand{\en}{\end{enumerate}}
\newcommand{\bc}{\begin{center}}
\newcommand{\ec}{\end{center}}

\def\epem{\ifmmode{e^+ e^-} \else{$e^+ e^-$} \fi}

\newcommand{\Dir}{\kern -6.4pt\Big{/}}
\newcommand{\Dirin}{\kern -10.4pt\Big{/}\kern 4.4pt}
\newcommand{\DDir}{\kern -7.6pt\Big{/}}
\newcommand{\DGir}{\kern -6.0pt\Big{/}}

\title{Higgs boson self-couplings at the LHC as a probe 
       of extended Higgs sectors}

\author{M. Moretti \\ 
Dipartimento di Fisica -- Universit\`a di Ferrara and
INFN -- Sezione di Ferrara, Via Paradiso 12, 44100 Ferrara, Italy}
\author{S. Moretti \\
School of Physics \& Astronomy, University of Southampton,
Highfield, Southampton SO17 1BJ, UK. E-mail: {stefano@hep.phys.soton.ac.uk}}
\author{F. Piccinini \\
INFN -- Sezione di Pavia and Dipartimento di Fisica Nucleare 
e Teorica, via Bassi 6, 27100 Pavia, Italy}
\author{\thanks{On leave of absence 
        from: Dipartimento di Fisica Teorica, Torino and INFN 
        Sezione di Torino, Italy}
R. Pittau \\
Departamento de F\'{\i}sica Te\'orica y del Cosmos
and Centro Andaluz de  F\'{\i}sica de Part\'{\i}culas 
Elementares (CAFPE), Universidad de Granada, E-18071 Granada, Spain}
\author{\thanks{Present address: Centro Studi 
e Ricerche ``E.~Fermi'', via Panisperna 89/A, 00184 Roma, Italy}
A.D. Polosa \\
LAPTH, BP 110, Chemin de Bellevue, 74941 Annecy-le-Vieux Cedex, France}

\abstract{A study of two-Higgs-doublet models (2HDM) in the
decoupling limit reveals the existence of parameter configurations
with a large triple-Higgs self-coupling as the only low-energy trace of the
departure from a Standard Model (SM) Higgs sector. This observation
encourages attempts to search for double Higgs production at the Large
Hadron Collider (LHC) even in mass regions which have been shown to be
very hard to probe in the context of SM-like Higgs self-couplings. In
this document we consider the case of an Intermediate Mass Higgs (IMH)
boson, with 120~GeV~$\lsim~M_H~\lsim$~140 GeV, produced in pairs via
vector-vector fusion, Higgs-strahlung and associate production with
heavy-quarks and decaying into $b\bar b$ pairs.  After a detailed
signal-to-background analysis, we confirm that the observation of a
Higgs-pair signal is very challenging in the framework of the SM, even at
the LHC with upgraded luminosity. In contrast, we verify that the 
sensitivity is
sufficient to detect departures from the SM which would escape
detection in the measurements of the single-Higgs production channels.}
\preprint{FNT-T/2004/03, SHEP-03-12, CAFPE-26/03 \\ 
          UG-FT-156/03, LAPTH-1028/04, BA-TH/495/04}

\keywords{Electro-Weak Symmetry Breaking, 
Higgs potential, 
Hadron Colliders, 
Beyond the Standard Model}

\begin{document}

\section{Introduction}
\label{sec:intro}
The ability of the ATLAS and CMS experiments at the LHC to detect a SM
Higgs boson over the full range of allowed masses has now been
established by several detailed studies~\cite{AtlasCMS}.  According to
the latter, a
multitude of production and decay channels will be accessible and
will enable the determination of several of the couplings
of the Higgs boson with accuracies that, depending on the precise
value of the Higgs mass, $M_H$, can be as good as 10\%~\cite{AtlasCMS}.
Possible significant departures from the SM expectations will allow to infer
that the underlying Higgs sector is not as simple as postulated in the
SM. However, there is no guarantee that non-SM Higgs sectors will become
manifest via these measurements. Examples are given by 
2HDMs, including Supersymmetric
models, where the spectrum of Higgs
bosons beyond the lightest one could be very heavy and the couplings
of the latter to fermions and gauge bosons reduce to those of the SM.  We
explore in this paper the possibility that in this limit a large
deviation from the SM value $\lambda_{HHH}^{(0)}$
of the triple-Higgs self-coupling involving the lightest Higgs state,
hereafter $\lambda_{HHH}$, is allowed to survive. Such a circumstance would
select the production of Higgs boson pairs as a possible channel for
the identification of a non-SM Higgs structure.

In Section~\ref{sec:SMHH} we review known results about Higgs-pair
production in the SM.  In Section~\ref{sec:2hdm} we discuss the main
features of the decoupling limit in 2HDMs and demonstrate the
existence of parameter sets leading to strongly enhanced triple-Higgs
self-couplings, while maintaining the deviations of the couplings to
fermions and to gauge bosons 
below the sensitivity range of direct LHC measurements. We
verified that for (some of) these models the analysis of Higgs-pair production
can be performed rather reliably (for the channels we consider in the
analysis) by simply rescaling in the Lagrangian the
value of \lambdahhhzero, keeping all other Higgs couplings fixed to their
SM values\footnote{For such channels this is due to
  the fact that, if we isolate the diagram which involves the triple-Higgs
  self-coupling,  its contribution is a constant at high
  energies. Therefore, despite the fact that
  $\lambdahhh\ne\lambdahhhzero$ breaks gauge invariance, 
  it is not crucial to enforce the unitarity of the cross section and
  it makes sense to assess the impact of changing the
  $\lambdahhh$ value while keeping all
  other couplings fixed. We have explicitly verified, as discussed later, 
  that this is indeed the case in the context of the 2HDM, 
  namely, with a gauge invariant calculation.}.
 Notice that this implies that our analysis is model
independent\footnote{In this spirit, we
will use the label $H$ to identify a light Higgs state
in both the SM and a generic extension, while
reserving the symbol $h$ specifically to the lightest Higgs boson of our
example scenario, i.e., the
2HDM discussed in Section~\ref{sec:2hdm}.}: 
our conclusions apply to any extended
Higgs sector in regions of parameter space where the only observable
departure from the SM is a sizeable deviation of \lambdahhh\  
from \lambdahhhzero.  
In Section~\ref{sec:bgs} we focus on
the study of Higgs-pair production at the LHC for 
IMH  bosons, namely with 
120~GeV$\lsim M_H\lsim $140~GeV. In this
region the detection of SM Higgs pairs is particularly challenging,
due to the very large QCD backgrounds and the very small signal 
cross sections. We extend previous analyses by
considering the production channels induced by
vector-vector fusion,
Higgs-strahlung and
associated production with a top quark pair.  For all
these channels we assume the $H\to b\bar{b}$ decay mode and we
perform a background study. In Section~\ref{sec:results} we present
our assessment of the prospects to constrain deviations from the SM
expectations, both at the LHC and at the so-called SLHC \cite{SLHC},
the tenfold luminosity increase option of the
LHC. Section~\ref{sec:summa} contains our summary and conclusions.

\section{Higgs-pair production at the LHC}
\label{sec:SMHH}
The complete reconstruction of the Higgs potential necessarily requires
the measurement of the Higgs self-couplings. These include a
trilinear and a quartic interaction, parameterized by the coupling
constants \lambdahhh\ and \lambdahhhh, which in the SM take the
following values:
\begin{equation}\label{lambdas}
\lambdahhhzero=-3\, \frac{M_H^2}{v} \; , \quad \lambdahhhhzero=-3 \,
\frac{M_H^2}{v^2}, 
\end{equation}
where $v=246$ GeV is the Higgs doublet vacuum expectation value. 
 A direct measurement of
\lambdahhh\ can be obtained via the detection of Higgs boson
pairs, wherein a contribution is expected from the production of a
single off-shell Higgs boson splitting into two.  However, the
corresponding graphs will always be accompanied by diagrams where the
two Higgses are radiated independently, with strength proportional to
the Yukawa or gauge couplings. As a result, different production
mechanisms will lead to different sensitivities of the Higgs-pair
production rates on the value of \lambdahhh\  (notice that
$\lambdahhh$ contributes, through loop radiative corrections, 
also to vector-boson propagators and vertices \cite{vdbji},
giving however
effects that are too tiny to be detected even in precision observables).

The leading production channels of Higgs boson pairs 
at hadron colliders \cite{ggHH}--\cite{VHH}
(see also \cite{review}--\cite{direct}) are basically the
same used for single-Higgs boson searches, namely:
\bea\label{procs}
gg &\to& {HH} ~ (gg{\mbox{~fusion}}),~~gg,q\bar q\to Q\bar Q HH ~
({\mbox{heavy-quark~associated~production}}), \nn \\ 
q q^{(')}&\to& q q^{(')} {HH} ~ ({\mbox{vector-vector~fusion}}),
~~q\bar q^{(')} \to V {HH} ~ ({\mbox{Higgs-strahlung}}),
\eea
wherein two Higgs bosons instead of one enter the final 
states and with
$V=W^\pm~{\rm or}~Z$, $Q=b,t$ and $q^{(')}$ referring to any possible
light (anti)quark flavor combinations\footnote{The case $Q=b$ in
heavy-quark associated production is actually irrelevant in the SM.
In a general 2HDM it may become significant. However, we will be interested 
only in the decoupling regime, which will be specified in more detail 
in the following Section. In this limit the departure of the cross section 
from the SM value is contrained to be at most within $70$\%. 
Hence, we have ignored the case $Q=b$ altogether in our study.}. 
The corresponding production rates 
are shown in Fig.~\ref{fig:sigHH} (borrowed from~\cite{review}),
where we have superimposed those for heavy-quark associated production. 
The arrows
indicate the variation in rate expected when changing the Higgs trilinear
self-coupling in the range $\lambdahhhzero/2 < \lambdahhh <
3/2 \, \lambdahhhzero$. 
Depending on the value of \mh, different decay channels
dominate the final state \cite{direct}. 
For a so-called IMH boson, with  120 GeV $\lsim\mh\lsim 140~\gev$, the decay
phenomenology is dominated by the channel
$H\to~b\bar{b}$. For $\mh\gsim140~\gev$, the  $H\to W^{\pm (*)}W^\mp$ 
and $H\to Z^{(*)}Z$ modes share the largest
fractions of the decay rate. 

Given the rather low Higgs-pair
production cross sections and the potentially large
backgrounds associated with final states with the best decay rates
(i.e., involving four $b$-quarks), naive arguments 
lead to the expectation that detection of IMH boson pairs via
$gg\to HH\to b\bar b b\bar b$  within the SM 
is most probably not feasible at the LHC and 
very difficult at the SLHC  \cite{SLHC}. This had been suggested already
in Refs.~\cite{Remi,LesHouches} (see also Ref.~\cite{standard}) and later
quantitatively confirmed in Ref.~\cite{Baur1}, where the production and 
decay channel $gg\to HH\to b\bar b \tau^+\tau^-$ was shown to provide
a  better sensitivity to \lambdahhh.  The potential of the rarer signatures 
$gg\to HH\to b\bar{b}\gamma\gamma$ and $gg\to HH\to b\bar{b}\mu^+\mu^-$ has
been reviewed in Ref.~\cite{hep-ph/0310056}, where the
former was found to provide at the SLHC some  
limited scope in constraining \lambdahhh\ when $M_H\approx120$ GeV. 
For a heavier Higgs boson, $M_H\gsim150$ GeV,
the situation is much brighter in comparison. In Ref.~\cite{Baur2}
it was found that
it should be possible at the LHC with design luminosity to establish that 
the SM Higgs boson has a non-zero trilinear self-coupling and that the ratio
$\lambdahhh / \lambdahhhzero$ can be restricted to the range 
0--3.8 at 95\% confidence level (CL), by exploiting  
$gg\to HH$ production and decay via $HH\to W^{\pm (*)}W^\mp
W^{\pm (*)}W^\mp$, in a variety of leptonic and/or hadronic final states.
At the SLHC~\cite{SLHC}, such limits can be improved further and even
4--5$\sigma$ excesses can be established.

All such studies were based on the leading production channel of
SM Higgs bosons, namely gluon-gluon fusion (see Fig.~\ref{fig:sigHH}).
We present here the first results of studies performed
 in the case of IMH boson pairs produced via the other three production 
modes in Eq.~(\ref{procs}). The reason to exploit the
latter is due to the additional triggers available
in each case, with respect to the $gg\to HH$ mode:
forward/backward jets in vector-vector fusion
and high transverse-momentum
leptons/light jets from gauge boson decays in both
heavy-quark associated production and Higgs-strahlung.
This could in principle enable one to perform additional 
independent measurements of \lambdahhh\ for an IMH boson, for
which the current situation is extremely problematic. Although we
will see that, as a result of 
our parton-level study, the available statistics
is in general too low for quantitative estimates
of \lambdahhhzero, we will show that there is room for a signal, or
for non-trivial limits, in the context of the 2HDM discussed in the
following Section.

\EPSFIGURE[!t]{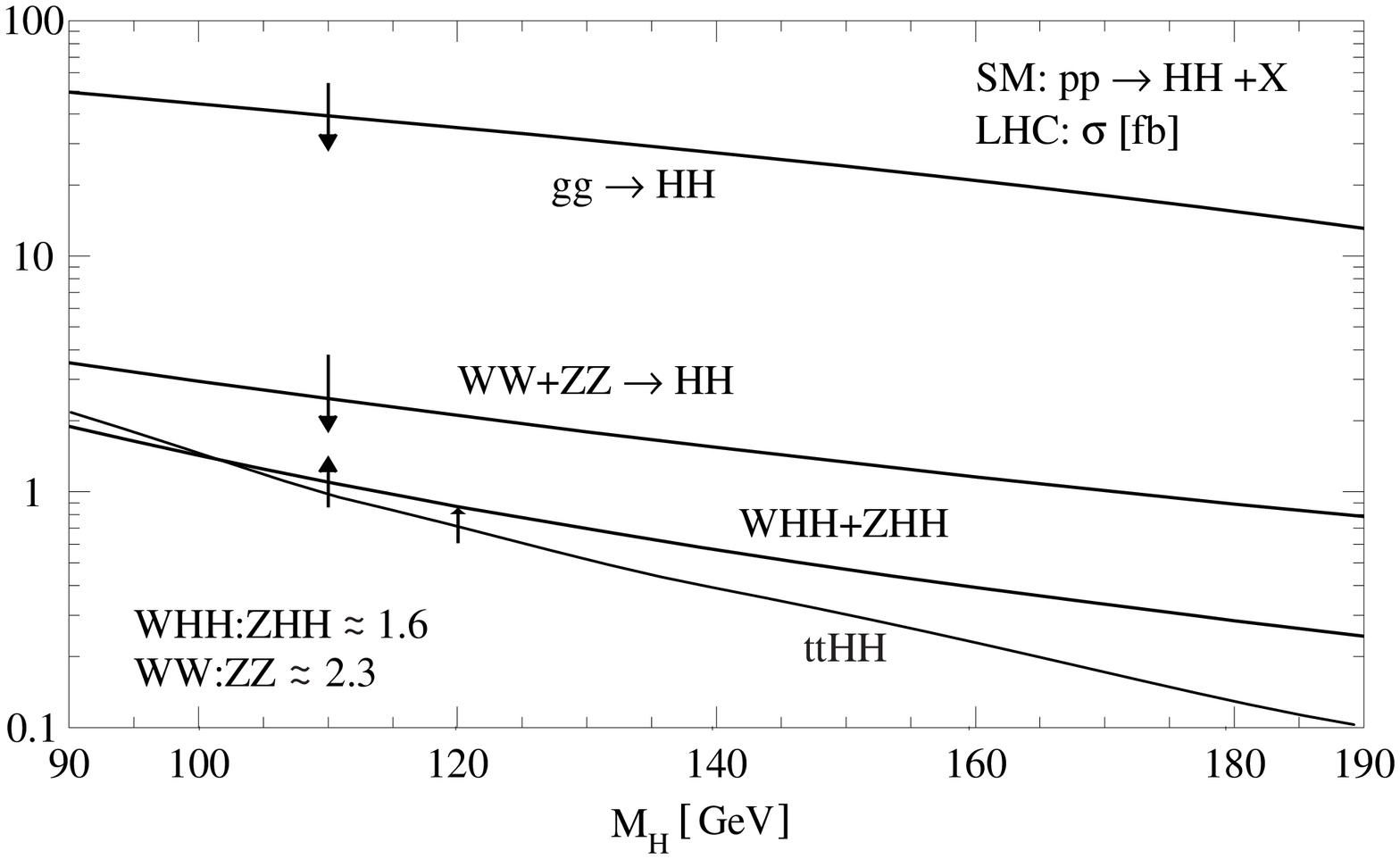,height=10cm,width=11cm}{Cross sections 
for Higgs-pair production in the SM
via gluon-gluon fusion, vector-vector fusion and 
Higgs-strahlung (from~\cite{review}), plus top-quark associated
production.  The vertical
arrows correspond to a variation of $\lambdahhh$ from 1/2 to 3/2 of
the SM value.
\label{fig:sigHH}}

\section{The decoupling limit of 2HDMs}
\label{sec:2hdm}
A study of the decoupling limit of 2HDMs has recently been 
presented in~\cite{2HDM} (see also \cite{Fawzi}), where 
the general expressions for the
spectrum and couplings of a generic, non-CP violating, 2HDM are
derived.  
In particular, it is shown (see Eqs.~(103)--(106) in~\cite{2HDM}) 
that the approach to
decoupling of the $hVV$, $ht\bar t$\footnote{The expression
for the case of $b$-quarks is obtained from that for $t$-quarks
upon the replacement $\cot\beta\leftrightarrow-\tan\beta$.} and $hhh$ vertices
of the lightest 2HDM Higgs state $h$ (of mass $m_h$)
can be parameterized as follows:
\ba
\frac{g^2_{hVV}}{g^{(0)}_{HVV}\,^2} &\sim &
  1-\frac{\hat{\lambda}^2v^4}{m_A^4} \; , \\
\frac{g^2_{htt}}{g^{(0)}_{Htt}\,^2} &\sim &
  1+\frac{2\hat{\lambda}v^2 \, \cot\beta}{m_A^2} \; , \\
\frac{\lambda_{hhh}^2}{\lambdahhhzero\,^2} &\sim &
  1-\frac{6\hat{\lambda}^2v^2 }{\lambda m_A^2}\; ,
\ea
where the suffices $H$ and $(0)$ label the SM quantities, 
$m_A$ is the mass of
the CP-odd Higgs boson, $\tan\beta$ is the ratio of expectation values
of the up-type and down-type Higgs doublets, $\lambda$ and $\hat\lambda$
are dimensionless parameters of the 2HDM. 
Notice that the deviations from the
decoupling limit are proportional to $\epsilon=\hat\lambda v^2/m_A^2$
and $\epsilon^2$ in the case of the couplings to fermions and gauge bosons 
whereas the self-coupling is proportional to $\epsilon\hat\lambda/\lambda$. 
The possibility that the ratio $\hat\lambda/\lambda$ be large 
allows for the triple-Higgs self-coupling to remain large even
when the other couplings are converging to their SM values. We analyzed
this possibility by implementing the exact couplings of a generic 2HDM
as discussed in~\cite{2HDM} and scanning the parameter space in the
range $1<\tan\beta<50$\footnote{Note that in
the scans we have traded the Higgs mixing angle
$\alpha$ for the lightest Higgs boson mass, $M_H$,
as independent 2HDM input parameter.}, 
$-4\pi < \lambda_i < 4 \pi$ for all couplings
$\lambda_i$ defined in~\cite{2HDM},
$i=1,...7$. 
While in the case $\lambda_{6,7}=0$ 
full analytical constraints can be derived 
guaranteeing the Higgs potential be bounded~\cite{2HDM},  
in the general scenario, 
$\lambda_{6,7}\neq 0$, the stability of the potential 
has to be checked numerically for every point considered in the 
parameter space. Our general scan was subject to the constraints of 
tree-level unitarity~\cite{2HDM} and to the requirement that the 
couplings $g^2_{hVV}$, $g^2_{htt}$ and $g^2_{hbb}$ differ from the SM 
values by no more than 30\%, 30\% and 70\%, respectively. These values 
reflect the measurement accuracies expected after 300 \ifb\ 
of accumulated LHC luminosity~\cite{AtlasCMS}\footnote{Also a more optimistic 
scenario of 20\%, 20\% and 30\% of measurement 
accuracies~\cite{leshouches03} has been investigated, yielding
conclusions similar to the ones outlined below, though over a restricted
parameter range (see also next footnote). }. 
The distribution of $r=\lambda_{HHH}/\lambda_{HHH}^{(0)}\equiv
\lambda_{hhh}/\lambda_{HHH}^{(0)}$ for the three Higgs mass
values of 120, 130 and 140 GeV in the general case ($\lambda_{6,7}\neq 0$)
is shown in Fig.~\ref{fig:ldist}, where the scan assumed 
equiprobable input values for all 2HDM inputs. Some regions of the 2HDM parameter space 
can result in small values for $m_H$ and $m_A$. Hence, we have imposed a 
further constraint of no visibility 
at $3\sigma$ level of the CP-even $H$-state at the LHC through 
the dominant $g g$ fusion production channel, exploiting the projected 
significance of ATLAS and CMS for SM Higgs production. 
The $gg\to H$ production cross section 
has been estimated isolating the $b$- and $t$- loop contributions in the 
{HIGLU} \cite{higlu} SM predictions and rescaling 
them with the proper 2HDM Higgs-quark couplings. 
As a result of this requirement, 
all the allowed points in Fig.~\ref{fig:ldist} 
correspond to $m_H \gsim 300$~GeV and $m_A \gsim 250$~GeV. 
(In the decoupling limit the dynamical behavior of the $A$ state 
is very similar to that of $H$, so that one can safely assume 
that whenever the latter is undetectable the former is too.) 
From now on we shall refer to the region of parameters which survive 
the above constraints as {\em decoupling limit} of the 2HDM.
The scan of all $\lambda_i$ leads to models with values of
$r$ in the ranges\footnote{In the 
case of the more optimistic uncertainty scenario we get 
$-3.5\lsim r\lsim 18$, 
$-3\lsim r\lsim 17$ and 
$-2\lsim r\lsim 16$ for $M_H = 120$, $130$ and $140$~GeV, respectively. 
Notice that we did not include so far constraints on $g^2_{h\tau\tau}$, 
 because 
a full detector simulation for this case is still lacking, unlike
the other cases. However, for completeness, we quote here some numbers 
also in presence of $g^2_{h\tau\tau}$ limits, though we will
not adopt them in the remainder of the analysis. 
With a projected 10\% accuracy on the latter, hence assuming 
30, 30 and 10\% as constraints in the scan (with the error 
on $g^2_{h\tau\tau}$ 
replacing the one on $g^2_{hbb}$), the above numbers change to
$-1.5 \lsim r\lsim 18.6$, 
$-1.6 \lsim r\lsim 34.2$ and 
$-1.2 \lsim r\lsim 12.5$ for $M_H = 120$, $130$ and $140$~GeV, respectively. 
Assuming 20, 20 and 10\% accuracies, the most optimistic case of all, we have 
$-1.5 \lsim r\lsim 4.1$, 
$-1.6 \lsim r\lsim 12.4$ and 
$-1.2 \lsim r\lsim 12.1$ for $M_H = 120$, $130$ and $140$~GeV, 
respectively. 
Hence, as it will be clear from the following kinematical analysis,
whichever the final outcome of the LHC analyses of single-Higgs production
channels, all described scenarios for $r$
will still be testable via double-Higgs
production.} 
\begin{eqnarray}
&& -8\lsim r\lsim 36, \, \, \, \, \, M_H = 120~{\rm GeV}, \nonumber \\
&& -7\lsim r\lsim 35, \, \, \, \, \,  M_H = 130~{\rm GeV},  \\
&& -6\lsim r\lsim 34, \, \, \, \, \,  M_H = 140~{\rm GeV}. \nonumber 
\label{decoup}
\end{eqnarray}

\EPSFIGURE[!t]{finldist.ps,height=10cm,angle=90}{\label{fig:ldist} Distribution
of the values of 
$r=\lambdahhh/\lambdahhhzero\equiv\lambda_{hhh}/\lambdahhhzero$ 
in the scans of the 2HDM parameters space defined in the text, 
for three values of the lightest Higgs boson mass. Normalization
is to unity.}

\section{Triple-Higgs self-couplings at the LHC}
\label{sec:bgs}
This Section describes the general setup of
our parton-level studies of signal and
background rates.  The numerical results are obtained by setting the
renormalization and factorization scales to $2M_{H}$ for the
vector-vector fusion and Higgs-strahlung signals and to
$\sqrt{\hat{s}}$ for heavy-quark associated production, with the
latter choice accounting for the effect of the large top mass. For the
various QCD backgrounds, the factorization/renormalization scale was
the average jet $E_T$ for the production of light jets and
$\sqrt{\hat{s}}$ for final states involving top quarks.  Both Higgs
processes and QCD backgrounds were estimated by using the parton
distribution function (PDF) set MRST99(COR01) \cite{pdfs}.  Both signal and
background estimates were based on exact tree-level matrix element
calculations using the \alpgen\ program \cite{ALPGEN}.  Furthermore,
all signal rates have been cross-checked by using independent
programs, based on either the HELAS subroutines \cite{HELAS} (for $q
q^{(')}\to q q^{(')} {HH}$ and $q \bar q^{(')}\to V {HH}$) or the
HELAC/PHEGAS package~\cite{Costas} (for $gg\to t\bar t HH$). As for
numerical input values of SM parameters, we adopt the \alpgen\
defaults. We assume that $b$-quark jets are distinguishable from
light-quark and gluon ones. Unless otherwise stated,
their tagging efficiency (including jet identification) is taken as $
\epsilon_{b}=50\% $ for each $ b $ with $ E_T^b>30~\gev $ and $ \vert
\eta^b \vert < 2.5 $ ($ \epsilon_{b}=0 $ otherwise) and
we also require all light- and $b$-jets to be separated,
$\Delta R_{bb,bj,jj} > 0.7$.  We do not assume $b$-jet charge
determination. Finally, in our parton-level analysis, we identify
jets with partons.

For the sake of definiteness, the signal results quoted in this
Section will refer to the case of SM values for \lambdahhh. The
analysis of the results in the context of our 2HDM example will be given in
the next Section.

\begin{figure}[!t]
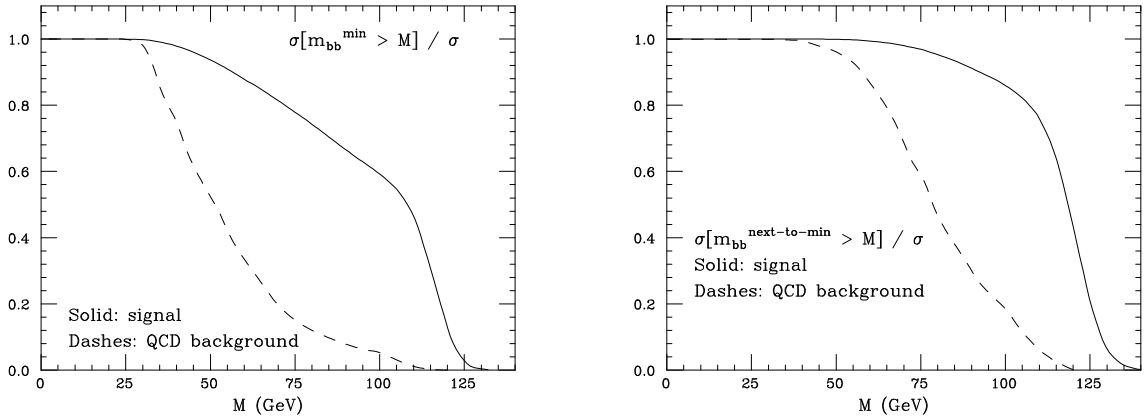

\begin{center}
\includegraphics[width=0.45\textwidth,clip]{mbbmin.eps} 
\hfill
\includegraphics[width=0.45\textwidth,clip]{mbbntmin.eps} 
\end{center}
\vskip -0.4cm 
\caption{\label{fig:mbbmin} Shapes of the integrated 
distributions of the minimum (left) and
next-to-minimum (right) $bb$ invariant masses in vector-vector fusion
and relative QCD background,
after the
application of the cuts discussed in the text 
up to Eq.~(\ref{eq:mhchi2}).
Both  signal and noise are shown for the case $ M_H=120$~GeV.}
\end{figure}

\subsection{{\it Vector-vector fusion}}
As seen in Fig.~\ref{fig:sigHH}, within the SM,
vector-vector fusion into Higgs boson
pairs is the subprocess displaying the second largest cross section
for any Higgs mass, behind $gg\to HH$, and is the most sensitive to 
deviations of the $\lambdahhh$ coupling from its SM value.
In this channel, before
any acceptance and selection cuts (and $b$-tagging efficiency) one
has production rates of about 1.4(0.7)[0.25]~fb 
for $M_H=120(130)[140]$ GeV. 
If one assumes efficient 
forward/backward jet tagging and high-purity sampling of
$b$-quarks, the dominant background is the QCD production of 
$b\bar{b}b\bar{b}jj$,
where the two extra jets are required to have $E_T^j>20~\gev$ and to go
one in the forward ($\eta^j>2.5$) and one in the backward ($\eta^j<-2.5$)
direction.  To enforce the reconstruction of the two Higgs
bosons, we require all four $b$'s in the event to be tagged 
and that at least one out of the three possible double 
pairings of $b$-jets satisfies the following mass constraint:
\be 
\label{eq:mhchi2}
(m_{b_1,b_2}-\mh)^2+(m_{b_3,b_4}-\mh)^2 < 2~\sigma_m^2, 
\ee 
where $\sigma_m= 0.12~ \mh$.
The $b\bar{b}b\bar{b}jj$ cross sections with this event selection are 
98(110)[130]~fb for $M_H=120(130)[140]$ GeV (with no $b$-tagging
efficiency). 
The QCD background is dominated by configurations
where the two $b$-quark pairs are produced by the splitting of recoiling
virtual gluons. These gluons tend to be off-shell by the smallest
possible amount, compatibly with the requirement of $\Delta R$
separation and minimum $\et$ of the $b$-jets.  The fake Higgs masses
are most often reconstructed by pairing $b$'s travelling back-to-back in
the transverse plane, namely pairs coming from the splitting of
different gluons. As a result, we expect that even for events where
two pairs exist passing the Higgs mass constraint, there will be
pairings of the four $b$'s with rather low invariant mass.
A very effective way to suppress the QCD background is therefore to cut 
on the minimum and and next-to-minimum $b \bar b$ invariant masses,
out of the six possible di-jet pairings, as can be appreciated
from Fig.~\ref{fig:mbbmin}.
The dependence of the signal and background cross sections on these 
additional constraints on $m_{bb}$ masses, also
including a more stringent cut on the $E_T$ of the forward/backward
jets, are displayed in Tab.~\ref{tab:sigcuts}. 
For our forthcoming numerical investigation on the 
sensitivity to anomalous triple-Higgs 
self-couplings we use the kinematic efficiency obtained with event selection 
c) of Tab.~\ref{tab:sigcuts}. 
The dependence of both signal and QCD
background on the transverse energy of the forward/backward jets
can be seen in Fig.~\ref{fig:pT-fwd-bwd}, e.g., after the cuts in a)
of Tab.~\ref{tab:sigcuts}.
Other potential background processes, such as 
$t \bar t b \bar b$, 
$t \bar t Z \to t \bar t b \bar b$, $t \bar t H \to t \bar t b \bar b$,
$j j Z Z \to j j b \bar b b \bar b$ and 
$j j b \bar b Z \to j j b \bar b b \bar b$, have been found to altogether have 
cross section lower than $10^{-1}$~fb for event selection c) 
of Tab.~\ref{tab:sigcuts} and have thus been neglected in the
remainder of our study. The same is true for the process 
$p p \to b \bar b + 4$~light jets, where two light jets are mistagged as 
$b$-jets 
with a fake efficiency of 0.01. 

Since in a realistic data analysis the background needs to be measured 
from the data, the question arises if the stringent cuts applied in the 
present analysis 
could produce peaking behaviours of the backgrounds in the signal 
region. 
To this aim we studied, for the most relevant background 
$p p \to 4b + 2$~forward jets, 
the handle offered by the distribution 
\begin{equation}\label{Mdist}
\frac{d \sigma}{d M}, \; \; {\rm with}\; \;   M = \sqrt{(m_{i1,i2} - M_H)^2 
+ (m_{i3,i4} - M_H)^2}
\end{equation}
and $\{i1,i2\}$, $\{i3,i4\}$ the two pairings minimizing $M$. For the 
case $M_H = 120$~GeV, the cut of Eq.~(\ref{eq:mhchi2}) implies 
a cut $M \lsim 20$~GeV. 
The distribution $d\sigma / dM$ (see Fig.~\ref{fig:syst}) does indeed 
exhibit a strongly peaked structure. However the peak of the distribution 
is outside the signal region at 120 GeV. 
We therefore expect that the background can be measured by 
looking at mass intervals away from the latter.  
As a cross-check of the accuracy of the extrapolation and as an assessment 
of the induced systematics, one can do the same exercise with a different 
nominal Higgs mass $\tilde M$, away from the signal region, for instance 
200~GeV (see also Fig.~\ref{fig:syst}). This 
allows also the investigation of the region $M \leq \tilde M \sqrt{2} \sigma$. 
 
Finally, concerning the trigger, in addition to the forward/backward jets, 
we have verified that more than 70\% of 
the signal events are characterized by at least one $b$-jet with 
$E_T > 100$~GeV, which could give an additional handle in selecting 
Higgs-pair production via vector-vector fusion. 

\TABULAR[!t]{|l|l|l|l|}{\hline
$M_H$~(GeV)   & $120$ & $130$ & $140$  \\ 
\hline 
Signal a) (fb)   & $0.085$ & $0.049$ & $0.021$ \\ 
Background a) (fb)  & $40$  & $35$   & $32$  \\
\hline
Signal b) (fb)   & $0.073$ & $0.043$ & $0.018$ \\ 
Background b) (fb)  & $21$  & $18$   & $17$  \\
\hline
Signal c) (fb)   & $0.036$ & $0.021$ & $0.008$ \\ 
Background c) (fb)  & $4$  & $3.5$   & $3$  \\
\hline}{The effects of additional cuts with respect to those
described in the text up to  
Eq.~(\ref{eq:mhchi2})
on the signal and background cross sections. 
Selection a) means $m_{bb}^{\rm min} >  50 ~\gev$ and 
$m_{bb}^{\rm next-to-min} >   100 ~\gev$, b) corresponds to 
$m_{bb}^{\rm min} >  70 ~\gev$ 
and $m_{bb}^{\rm next-to-min} >   110 ~\gev$ and c) to the latter
with the more stringent
cut $E_T^j > 40$~GeV (instead of $E_T^j > 20$~GeV) on the forward/backward 
jets. No $b$-tagging efficiency is included. 
\label{tab:sigcuts}}

\begin{figure}[!t]
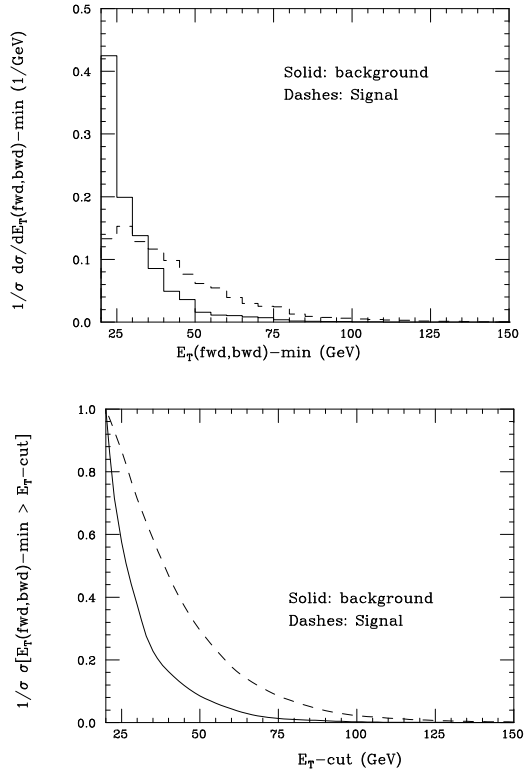

\begin{center}
\includegraphics[width=0.45\textwidth,clip,angle=0]{ptfwd.ps} 
\hfill
\vskip 0.5cm
\includegraphics[width=0.45\textwidth,clip,angle=0]{ptfwdint.ps} 
\end{center}
\vskip -0.4cm 
\caption{\label{fig:pT-fwd-bwd}
Shapes of the differential (up) and
integrated (down) distributions of the minimum
forward/backward jet transverse momentum 
in vector-vector fusion and relative QCD background, after the
application of the cuts discussed in the text up to 
Eq.~(\ref{eq:mhchi2})
Both  signal and noise are shown for the case $ M_H=120$~GeV.}
\end{figure}
\begin{figure}[!t]
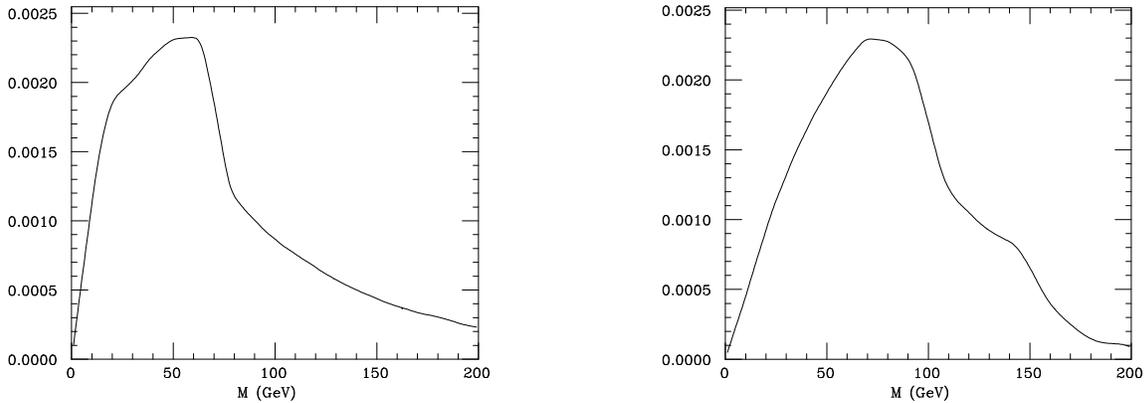

\begin{center}
\includegraphics[width=0.35\textwidth,clip,angle=90]{syst_120.eps} 
\hfill
\includegraphics[width=0.35\textwidth,clip,angle=90]{syst_200.eps} 
\end{center}
\vskip -0.4cm 
\caption{\label{fig:syst} 
Shape of the distribution of $M$ in Eq.~(\ref{Mdist})
 for the case $ M_H=120$~GeV (left) and $ M_H=200$~GeV (right).}
\end{figure}

\subsection{{\it Higgs-strahlung}}
In the Higgs-strahlung channel,  the cross section without any cuts
and tagging efficiencies, before the decays of the
$W^\pm$ and $Z$ bosons, 
but including the $HH\to b\bar b b\bar b$ decay rate,  is
0.41(0.19)[0.065]~fb 
for $M_H=120(130)[140]$ GeV. Let us first consider the case of 
leptonic decays of the gauge bosons, with $\ell=e,\mu$.
The cuts on $b$-jets
used here are the same as for the vector-vector fusion case,
with the exclusion of the constraints discussed in Tab.~\ref{tab:sigcuts}. 
$ZHH$ events are selected requiring  two charged leptons with 
$E_T^\ell>20$~GeV,  $|\eta^\ell|<2.5$ and invariant mass within 5~GeV
of the $Z$ mass. $W^\pm HH$ events are extracted
by isolating one charged lepton with 
$E_T^\ell>20$~GeV and  $|\eta^\ell|$ $<$ $2.5$ plus $E_T^{\rm{miss}}>20$~GeV.
The resulting (cumulative) signal amounts to 
0.0083(0.0048)[0.0019]~fb, for $M_H=120(130)[140]$ GeV,
prior to $b$-tagging efficiency.
The corresponding irreducible background arises from $V b {\bar b} b {\bar b}$
events, with a rate of about 0.5~fb for the above cuts, with very small 
Higgs mass dependence. We also considered  hadronic decays of both
$W^\pm$ and $Z$, yielding a $4b$ plus 2(light-)jets signature. Acceptance,
separation and mass cuts
on the $4b$ system are as usual, whereas we require the two light jets
to have $E_T^j>20~\gev$, $\vert \eta^j \vert $ $<$ $2.5$
and to satisfy 70~GeV~$<m_{{jj}}<$~100~GeV. The signal rates increase by 
a factor of about 4 with respect to the leptonic case, but the 
irreducible background increases by a factor of about 10. Moreover, here, 
one should also include the large contribution coming from QCD production of 
$jjb {\bar b} b {\bar b}$, rendering the channel completely hopeless.

\subsection{{\it Heavy-quark associate production}}
As intimated already, in the SM only the case 
$Q=t$ in Eq.~(\ref{procs}) has some phenomenological importance.
Ta\-king our u\-su\-al $HH\to b\bar b b\bar b$ signature leads
to final sta\-tes which include six $b$-quarks, two of them coming from
the decays of $t$ and $\bar t$. The largest backgrounds
arise from the QCD production of $ t\bar{t} b\bar{b} b\bar{b} $ and,
if the request of tagging were limited to four $b$-jets only, 
 $ t\bar{t} b\bar{b} jj $, where $ j $ represents a jet of gluons
 or light quarks. We estimated the $ t\bar{t} b\bar{b} jj $  background
 and obtained a rate of 1.5~pb, 
with usual jet acceptance cuts ($
 E_T^{b,j}>20$ $\gev$, $ \vert \eta^{b,j} \vert$ $<$ $2.5$) and
somewhat relaxed jet isolation requirements 
($ \Delta R_{bb,bj,jj}>0.4$). 
 This implies that tagging
 all six jets is man\-da\-to\-ry. 
With the above cuts, the 
$ t\bar{t} b\bar{b} b\bar{b} $ background is in fact only 3.6~fb.
 Notice that, for this channel, maintaining a high $
 b $-tagging efficiency is crucial, since it appears with the sixth
 power. If the SLHC luminosity led to a decrease from 50\% to 40\%,
 one would loose a factor of 4 in rate, and most of the advantage of
 the high luminosity gain would be lost. Notice also that in order to
 keep the $ t\bar{t} b\bar{b} jj $ background below the level of 
 $ t\bar{t} b\bar{b} b\bar{b} $ requires the fake tagging rate to be
 kept below 5\% for each jet. 
 To complete the reduction of the continuum background, we require
 that at least one out of all pairings of $ b $-quarks satisfies
 the constraint given in Eq.~(\ref{eq:mhchi2}). 
We consider a semi-leptonic signature with one $W^\pm$ decaying to leptons 
($e$, $\mu$ and their neutrinos) and the other one hadronically. 
So the final states are of the kind 
$b {\bar b} b {\bar b} b {\bar b} j j \ell \nu_\ell$. 
Events are selected imposing 
$E_T^j>20$~GeV,  $|\eta^j|$ $<$ $2.5$, $\Delta R_{jb,jj}>0.4$. 
We do not impose $E_T$, $\eta$ or
$\Delta R_{\ell j,\ell b}$ criteria 
on leptons from $W^\pm$ decay. 
The pre-selection criteria for $b$-quarks are  
$E_T^b>30$~GeV,  $|\eta^j|$ $<$ $2.5$, $\Delta R_{bb}>0.7$. 
With the above cuts, we obtain the signal 
cross sections $0.021$, $0.010$ and $0.0035$~fb (without including the 
branching ratio for the $t \bar t$ semi-leptonic signature, 8/27, but 
including the ${\rm BR}(H\to b \bar b)^2$), corresponding to 
$M_H = 120$, $130$ and $140$~GeV, respectively.
The irreducible background 
$t {\bar t} b {\bar b} b {\bar b}$ 
gives a cross section of 0.007~fb for any relevant 
Higgs mass. 
Both signal and background are expected to be 
sensitive to the choice of factorization/normalization
scale, as they originate from QCD induced processes, primarily
via gluon PDFs. Hence, 
we have checked the dependence of these numbers against variations of the 
factorization/normalization scale. For example,
if $Q^2 = (2 m_t + 2 M_H)^2$, we obtain instead 
the following signal cross sections: $0.025$, $0.012$ and $0.004$~fb 
for $M_H = 120$, $130$ and $140$~GeV, respectively, while
the corresponding background rate grows up to 0.013~fb. 
Detailed studies will be needed to assess to which extent a
control data sample can be identified to fix the overall background
normalization. 
The extraction of the signal from backgrounds could be feasible at 
a $1\sigma$ statistical level, but the number of events is too small 
even to see a signal event, as the cross sections quoted above do not include 
the $6b$-tagging  efficiency factor, i.e., $\epsilon_b^6$.  
Assuming {5$b$-tagging} instead, always 
with $\epsilon_b = 0.5$, the {signal events 
grow by a factor of } {7}, at the price of including further
background processes: namely,
$t {\bar t} b {\bar b}$~+~light jets.
With the usual fake tagging efficiency of 1\%, we checked that there is no 
sensible gain in statistical significance.

\section{Discussion of the results}
\label{sec:results}
As was clear from the signal and background rates obtained in the
previous Section, even the SLHC
luminosity option will not be sufficient to establish a statistically
significant signal for $HH$ production within the SM in the 
$4b$ channel. In this Section
we therefore consider the effects of an anomalous \lambdahhh\
coupling, in the
ranges obtained in Section~\ref{sec:2hdm}.
In order to do so, we
re-evaluated the signal rates in the presence of
non-SM triple-Higgs self-couplings by simply rescaling the value of
$\lambdahhhzero$ to a generic $ \lambdahhh$. The corresponding rates,
as a function of $\lambdahhh/\lambdahhhzero$, are shown in
Fig.~\ref{fig:lambdas}. Over the range allowed by our scan and by
unitarity, cross section enhancements by up to two orders of magnitude
can be obtained.
Exclusion limits (at $95\%$ CL) and signal evidence (at $3\sigma$) for
anomalous triple-Higgs self-couplings by combining all channels are found
in Tab.~\ref{tab:signcombined}. In Tab.~\ref{tab:signcombined2} the 
limits 
have been derived taking into account also a forward/backward jet 
recontruction efficiency of 80\%.
The by far most sensitive mode is vector-vector fusion, 
with heavy-quark associated production being of some relevance
only for small Higgs masses and positive 
$\lambda_{HHH}/\lambda_{HHH}^{(0)}$ values, as can 
be expected by looking at Fig.~\ref{fig:lambdas}. From the
same plot, it is also obvious why
the limits obtained via the Higgs-strahlung mode 
are always much weaker than those extracted in the other two channels. 

\EPSFIGURE[!t]{finlambdas.ps,width=0.7\textwidth,angle=90}{Dependence of 
the cross sections for the three processes 
$q q^{(')}\to q q^{(')} {HH}$, $gg,q\bar{q}\to t\bar t HH$ and
$q\bar{q}^{(')} \to V {HH}$ on $r=\lambdahhh/\lambdahhhzero$, 
in correspondence of three values of the Higgs mass, 
for the model setup discussed in the text.
\label{fig:lambdas}}

Notice that the described rescaling is not a gauge invariant 
operation, because diagrams involving other genuine 2HDM fields (namely,
$H, A$ and $H^\pm$)
in the Lagrangian, 
as well the rescaling of the Higgs-to-SM-particle couplings to the 2HDM values,
are neglected. Therefore, one might suspect that the cross sections
in Fig.~\ref{fig:lambdas} are anomalously enhanced by unitarity 
violation effects.
To investigate this possibility we have re-computed the 
vector-vector fusion cross section, including 
the full set of 2HDM diagrams,
for several hundreds of points uniformly distributed 
over the 2HDM parameter space,
fulfilling the decoupling conditions of Section~\ref{sec:2hdm} and  
with $r$ in the ranges accessible at the SLHC, namely
$r< -2.3(-3.1)[-5.3]$ or $r>3.9(5.9)[8.3]$, in correspondence
of $M_H=120(130)[140]$ GeV.

As a results of this exercise, we did find points, for any value of $r$, 
where the 2HDM cross section agrees almost exactly with the one depicted 
in Fig.~\ref{fig:lambdas}.
Since the calculation is now gauge invariant, this fact alone 
demonstrates that the approximation implicit in Fig.~\ref{fig:lambdas}
(namely neglecting additional graphs and coupling modifications) does not
lead to an artificial enhancement of the sensitivity to $\lambda_{HHH}$.
Furthermore, we found a large fraction ($\approx70 \%$) 
of points where the cross section is reproduced by the above described
approximation
within a factor two. Finally, for a small, but sizeable, portion of points
the cross section is substantially underestimated while for even fewer
points is grossly overestimated.
For completeness, we also tried to understand the origin of 
such discrepancies. Firstly, the most noticeable ones are due to the 
BR$(H \to b\bar b)$, as our definition of decoupling region 
allows for rather sizeable deviations of the latter
 from the SM value and, since
we are looking for two Higgs bosons decaying into $b \bar b$, 
this BR enters quadratically into our predictions. Secondly, also the
diagrams  proportional to 
$g_{hVV}$ ($V= W,Z$) can give an important contribution: 
although the decoupling limit strongly constraints deviations of $g_{hVV}$' 
from the SM value, the consequences of the latter
can be enhanced by large destructive interferences. 
Once these two effects are accounted for, the overall agreement is fairly good:
the contribution of the SM diagrams to the total cross section is correct 
within a 20\% accuracy, except for a small fraction of points where the
production rate is strongly underestimated.
We have finally
verified that the latter effect is due to neglecting diagrams where 
a light 2HDM Higgs pair is produced via the decay of 
the heavy 2HDM Higgs boson, $H\to hh$.
Notice that the effect of this additional contribution reinforces our conclusion that 
the signal is detectable, actually suggesting that a search is possible 
even in regions where $\lambda_{HHH}$ exhibits deviations from 
the SM value smaller than those appearing in 
Tabs.~\ref{tab:signcombined}--\ref{tab:signcombined2}.


\TABULAR[!t]{|l|rr|rr|rr|}{\hline
\mh~(GeV) & $120$ & & $130$ & & $140$ & \\ 
\hline
LHC, 95\% CL & $-4.8$ & $7.5$ & $-6.0$ & $9.0$ & $-9.5$ & $12.4$  \\ 
SLHC, 95\% CL & $-1.8$& $3.7$ & $-2.5$ & $5.3$ & $-4.4$ & $7.4$  \\    %
\hline
LHC, 3$\sigma$  & $-6.6$ & $9.3$ & $-8.1$ & $11.0$ & $-12.4$ & $15.4$  \\ 
SLHC, 3$\sigma$ & $-2.7$ & $5.1$ & $-3.6$ & $6.5$ & $-5.9$ & $8.9$  \\ %
\hline
}{\label{tab:signcombined} Constraints on the ratio 
$\lambda_{HHH}/\lambda_{HHH}^{(0)}$ using all the 
channels described in Section \ref{sec:bgs}. (The results are almost
completely driven by vector-vector fusion.) 
In the top box, the two values in each entry
correspond to $r_{\rm min}$, $r_{\rm max}$, where $r<r_{\rm min}$ and
$r>r_{\rm max}$ define the range which can be excluded at 95\% CL 
(first row)    
or probed at the 3$\sigma$ level 
(second row),  
at both the LHC and SLHC). The number of signal events corresponding 
to $3 \sigma$ significance are about 120, 
110 and 100 for $m_h\equiv M_H = 120$, 
$130$ and $140$~GeV, respectively, at the SLHC for the event selection c) 
of Tab.~\ref{tab:sigcuts}, with a background of about 1500, 1300 and 
1100 events, respectively.}

\TABULAR[!t]{|l|rr|rr|rr|}{\hline
\mh~(GeV) & $120$ & & $130$ & & $140$ & \\ 
\hline
LHC, 95\% CL & $-5.6$ & $8.0$ & $-6.9$ & $9.8$ & $-10.8$ & $13.7$  \\ 
SLHC, 95\% CL & $-2.2$& $3.8$ & $-3.0$ & $5.7$ & $-5.1$ & $8.0$  \\    %
\hline
LHC, 3$\sigma$  & $-7.6$ & $10.0$ & $-9.2$ & $12.0$ & $-14.1$ & $16.9$  \\ 
SLHC, 3$\sigma$ & $-3.2$ & $5.3$ & $-4.2$ & $7.0$ & $-6.8$ & $9.7$  \\ %
\hline
}{\label{tab:signcombined2} The same as in Tab.~\ref{tab:signcombined}, 
including a $0.8$ reconstruction efficiency for each forward/backward 
jet.}

In closing, we note that our results are not confined
to the 2HDM in the decoupling limit but are truly model independent, thereby
being applicable to other, more exotic Higgs sectors displaying a similar 
decoupling 
behavior between the lightest CP-even Higgs state and the heavier ones.
The study of these models is however beyond the scope of this work.

\section{Summary and outlook}
\label{sec:summa}
In the SM, even at the SLHC, the measurement of $\lambdahhhzero$ 
for $M_H \leq 140$~GeV by using 
subleading Higgs-pair production channels, with the two Higgs particles
both decaying into $b\bar b$ pairs, appears to be problematic,
just like in the case of the leading mode studied in previous literature.  
However, by studying generic 2HDMs in the decoupling limit, we found large 
regions
of parameter space where the triple-Higgs self-coupling
$\lambdahhh$ can differ considerably from the SM value, while keeping 
all other Higgs interactions with fermions and gauge bosons experimentally 
consistent with the SM limits. For the allowed regions, 
where the full 2HDM result can be emulated in a model-independent way 
(within an uncertainty of a factor of at worst two) by
simply rescaling the SM triple-Higgs self-coupling to the 2HDM value,
we have analysed the detectability (or otherwise) of various processes 
of the type $p p \to X HH \to X 4b $ in the mass range $M_H \leq 140$~GeV
at the (S)LHC. In
particular, signals at the level of 3$\sigma$ from non-trivial regions of 
the 2HDM parameter space can be obtained already at the LHC, with the SLHC
extending the scope considerably further. 
The model-independent approximation used to perform 
the phenomenological analysis has been thoroughly tested against 
a complete 2HDM calculation for the case of the vector-vector fusion process, 
which is the most sensitive to variations of the $\lambda_{HHH}$ coupling. 
The approximation can be improved (in a model-dependent way, however) by 
introducing the modified 
${\rm BR}(H \to b \bar b)$ and coupling $g_{HVV}$  
($V=W,Z$) as predicted in the 
2HDM. The only remaining discrepancies can be 
accounted for, within an accuracy of the order of 30\%, by the contribution 
of the production of a heavy Higgs state and its decay into the light 
Higgs pair, being the contribution of the remaining non-SM diagrams negligible.

In summary, Higgs-pair production at the (S)LHC could be an 
important channel to unravel a non-standard Higgs sector. In particular, 
in the mass window $120 \lsim M_H \lsim 140$~GeV, the $4b X$ signature 
studied in this document 
will be the only accessible channel which could
show sizeable departures of the trilinear coupling
involving the lightest Higgs state from the SM value, 
in a scenario where the single-Higgs production channels do 
not show detectable anomalies with respect to the SM predictions. 

Finally, our further efforts  will concentrate 
on the case of triple-Higgs self-couplings among neutral 
Higgs bosons in a general 2HDM away from the decoupling limit
as well as in the Minimal Supersymmetric Standard Model (MSSM)
(see \cite{H0HpHm} for  2HDM and MSSM studies of similar 
vertices involving charged Higgs bosons).

\acknowledgments
We thank M.L.~Mangano for his participation in the early stages of this 
project and subsequently for many insightful discussions and comments. 
We also thank F.~Boudjema for valuable suggestions, 
C. Papadopoulos for contributing to the estimate of the heavy-quark 
associated production cross sections 
and M. Spira for useful information on the use of the {HIGLU} programme. 
SM finally thanks the UK-PPARC for financial support and
J. Rathsman for correspondence. 
ADP thanks the Physics Department of the University of Bari for its 
kind hospitality. 
MM and RP acknowledge the financial support of the European Union 
under contract HPRN-CT-2000-00149. RP also acknowledges the financial support 
of MIUR under contract 2001023713-006 and of MECD under contract
SAB2002-0207. FP wishes to thank the CERN Theory Unit for
partial financial support while part of this work was being carried out.


\begin{thebibliography}{99}

{\small

\bibitem{AtlasCMS} 
S. Asai et al.,
SN-ATLAS-2003-024; 
M. D\"uehrssen, ATL-PHYS-2003-030; 
S. Abdullin et al., CMS Note 2003/33.

\bibitem{SLHC} 
F. Gianotti, M.L. Mangano and T. Virdee (conveners),
{\tt hep-ph/0204087}.

\bibitem{vdbji} J.J.~van~der~Bij, Nucl.~Phys.~{\bf B267} (1986) 557.

\bibitem{ggHH} E.W.N.~Glover and J.J.~van der Bij, Nucl.\ Phys.\ {\bf
B309} (1988) 282; D.A. Dicus, C. Kao and S.S.D. Willenbrock,
Phys. Lett. {\bf B203} (1988) 457; G.~Jikia, {Nucl.~Phys.} {\bf B412}
(1994) 57; T. Plehn, M. Spira and P.M. Zerwas, Nucl. Phys. {\bf B479}
(1996) 46, Erratum, ibidem {\bf B531} (1998) 655.

\bibitem{VVHH} W.-Y. Keung, Mod. Phys. Lett. {\bf A10} (1987) 765;
O.J.P. \'Eboli, G.C. Marques, S.F. Novaes and A.A. Natale,
Phys. Lett. {\bf B197} (1987) 269; D.A.~Dicus, K.J.~Kallianpur and
S.S.D.~Willenbrock, Phys.\ Lett.\ {\bf B200} (1988) 187;
K.J.~Kallianpur, Phys.\ Lett.\ {\bf B215} (1988) 392; A.~Abbasabadi,
W.W.~Repko, D.A.~Dicus and R.~Vega, Phys.\ Rev.\ {\bf D38} (1988)
2770; Phys.\ Lett. {\bf B213} (1988) 386; A.~Dobrovolskaya and
V.~Novikov, Z.\ Phys.\ {\bf C52} (1991) 427.

\bibitem{VHH}
        V.~Barger, T.~Han and R.J.N.~Phillips, Phys.\ Rev.\ {\bf D38}
        (1988) 2766.


\bibitem{review} A. Djouadi, 
        W. Kilian, M. M\"uhlleitner and P.M. Zerwas,
        Eur. Phys. J. {\bf C10} (1999) 45. 

\bibitem{direct}
M. Spira, A. Djouadi, D. Graudenz and P.M. Zerwas, Nucl. Phys. 
{\bf B453} (1995) 17; 
Z.~Kunszt, S.~Moretti and W.J.~Stirling,
Z.\ Phys.\ {\bf C74} (1997) 479.

\bibitem{Remi} R. Lafaye, D.J. Miller, M. Muhlleitner and
 S. Moretti, 
{\tt  hep-ph/0002238}.

\bibitem{LesHouches}
A. Djouadi {et al.},
in 
{\tt hep-ph/0002258};
D. Cavalli
 {et al.},
in 
{\tt hep-ph/0203056}.


\bibitem{standard} E. Richter-Was {et al.}, Int. J. Mod Phys. {\bf
    A13} (1998) 1371; E. Richter-Was and D. Froidevaux, Z. Phys. {\bf
    C76} (1997) 665; J. Dai, J.F. Gunion and R. Vega, Phys. Lett. {\bf
    B371} (1996) 71, {ibidem}, {\bf B378} (1996) 801.

\bibitem{Baur1} U. Baur, T. Plehn and D. Rainwater,
Phys. Rev. {\bf D68} (2003) 033001. 

\bibitem{hep-ph/0310056} U. Baur, T. Plehn and D. Rainwater, 
Phys. Rev. {\bf D69} (2004) 053004.

\bibitem{Baur2} U. Baur, T. Plehn and D. Rainwater,
Phys. Rev. {\bf D67} (2003) 033003; Phys. Rev. Lett. {\bf 89} (2002) 151801.

\bibitem{2HDM} J.F. Gunion and H.E. Haber,
Phys. Rev. {\bf D67} (2003) 075019.

\bibitem{Fawzi} F. Boudjema and A.V. Semenov, Phys. Rev. {\bf D66}
(2002) 095007; I.F.~Ginzburg and M.~Krawczyk, {\tt hep-ph/0408011}; 
S.~Kanemura, Y.~Okada, E.~Senaha and C.-P.~Yuan, {\tt hep-ph/0408364}; 
S.~Kanemura, Y.~Okada, E.~Senaha, {\tt hep-ph/0410048}.


\bibitem{leshouches03} 
M. D\"uehrssen
et al., 
{\tt hep-ph/0406323}
and in 
{\tt hep-ph/0406152}.

\bibitem{higlu} M.~Spira, Nucl.~Instrum.~Meth. {\bf A389} (1997) 357.

\bibitem{pdfs} See: {\tt http://durpdg.dur.ac.uk/hepdata/pdf.html}.

\bibitem{ALPGEN} M.L. Mangano, M. Moretti, F. Piccinini, R. Pittau
and A.D. Polosa,
JHEP {\bf 0307}  (2003) 001.

\bibitem{HELAS} H.~Murayama, I.~Watanabe and K.~Hagiwara,
                       KEK Report 91--11, January 1992.

\bibitem{Costas} A. Kanaki, C.G. Papadopoulos,
Comput. Phys. Commun. {\bf 132} (2000) 306;
C.G. Papadopoulos, Comput. Phys. Commun. {\bf 137} (2001) 247.

\bibitem{theorem} T. Appelquist and J. Carazzone,
Phys. Rev. {\bf D11} (1975) 2856.
  
\bibitem{HDECAY} 
A.~Djouadi, J.~Kalinowski and M.~Spira,
Comput.\ Phys.\ Commun.\  {\bf 108} (1998) 56.

\bibitem{wjszk}
S.~Moretti and W.~J.~Stirling,
Phys.\ Lett. {\bf B347} (1995) 291,
Erratum, ibidem {\bf B366} (1996) 451.

\bibitem{H0HpHm} S. Moretti and J. Rathsman, Eur. Phys. J. {\bf C33} (2004) 41.

}
\end{thebibliography}
\end{document}